\documentstyle[12pt]{article}
\begin{document}
\begin{flushright}
CEBAF-TH-94-07 \\
\end{flushright}
\vskip .9cm
{\large\bf
 \centerline{Reduced phase space quantization of Ashtekar's gravity}
 \centerline{on de Sitter background}
}
\vskip .7cm
\centerline{Ignati Grigentch}
\begin{center}
{\it
 Department of Physics, Old Dominion University, Norfolk, VA 23529, USA \\
 and \\
  CEBAF Theory Group, MS 12H2 \\
 12000 Jefferson Avenue, Newport News, VA 23606, USA.
}
\end{center}
\vskip .6cm
\centerline{D.V.Vassilevich}
\begin{center}
  {\it  Department of Theoretical Physics, St. Petersburg University \\
198904, St. Petersburg, Russia
}
\end{center}
\begin{abstract}
We solve perturbative constraints and eliminate gauge freedom for Ashtekar's
gravity on de Sitter background. We show that the reduced phase space consists
of transverse, traceless, symmetric fluctuations of the triad and of
transverse,
traceless, symmetric fluctuations of the connection. A part of gauge freedom
corresponding to the conformal Killing vectors of the three-manifold can be
fixed only by imposing conditions on Lagrange multiplier. The reduced phase
space is equivalent to that of ADM gravity on the same background.
\end{abstract}

\quad PASC: \ 04.60.+n; \ 04.20.Fy
\newpage
\setcounter{page}{2}
\def\slash#1{#1 \hskip -0.5em / }

It is well known that quantum gravity is perturbatively non-renormalizable.
Certain hopes that it will be possible to obtain a non-perturbative
description of gravity are related to Ashtekar's variables $[1], [2]$ in which
the constraints become polynomial. The introducing of cosmological term in
this formalism was considered in $Ref. 3.$ Several papers were devoted to the
construction of the reduced phase space quantization in a framework of
Ashtekar's formulation. We mention recent ones $[4].$ Certain progress was
also achieved in the quantization of $2+1$ dimensional models (see for
example $Ref.5$ and references therein). And there are many other related
fields of research, as loop variables and minisuperspace models, which are
not addressed here.

Apart from the non-renormalizability there are such internal problems
arising in a framework of perturbative quantum gravity on de Sitter
background as the lack of uniqueness of one-loop predictions which
originates from ambiguous treatment of the zero-mode structure and
non-covariance of the path integral measure. Recently this difficulty was
resolved $[6]$ in a framework of the ADM gravity $[7].$ It is interesting,
however, to have a look into this problem from the point of view of another
canonical approach.

In this paper we analyse linearized constraints of Ashtekar's gravity
on de Sitter background and demonstrate that the complex reduced phase
space $[8]$ consists of symmetric, traceless, transverse fluctuations of
the densitized triad and transverse, traceless, symmetric fluctuations of the
connection. Transfer to the real phase space is done by imposing reality
conditions. It is demonstrated that a part of gauge freedom can be fixed
only by means of some additional conditions on Lagrange multiplier. This is
essential for the path integral quantization since this results in an
appearance of an additional Jacobian factor $[6].$ After this complete
fixation of gauge freedom the real reduced phase space is proved to be
equivalent to that of the ADM gravity. Our results generalize the results
for flat background space reported in $[2].$

This work can be considered as a starting point for quantization of full
non-linear theory in the case of non-trivial topology of space-time.

We begin with complex gravitational action in $3+1$ dimensions
\begin{equation}\label{eqn1}
S=\int d^4x\left (i{E_a}^i\partial_t{A_i}^a-N^aG_a-N^iG_i-NG_0\right )
\end{equation}
where, as usual, the densitized triad ${E_a}^i$ and connection ${A_i}^a$ are
the canonical variables;\ $G_a,G_i$ and $G_0$ are the Gauss law, the vector
and the scalar constraints respectively

\begin{equation}G_a={\cal D}_i{E_a}^i=0\end{equation}
\begin{equation}G_i={F_{ij}^a}{E_a}^j=0\end{equation}
\begin{equation}G_0=\varepsilon^{abc}{E_a}^i{E_b}^j{F_{ij}^c}-{\Lambda \over 3}
\varepsilon^{abc}\varepsilon_{ijk}{E_a}^i{E_b}^j{E_c}^k=0,\end{equation}

$\Lambda$ is the cosmological constants; $N^a, N^i$ and $N$ are Lagrange
multipliers, ${\cal D}_i$ is the covariant derivative with respect to the
connection ${A_i}^a$ , $F_{ij}^a$ is the field strength. We use $i,j,k,l\ldots$
to denote world indices, while $a,b,c,d\ldots$ are reserved for Lorentz
indices.

The reality conditions have the following form
\begin{equation}E=E^*,\ \quad A+A^*=2\Gamma\left (E\right ),\end{equation}
where $\Gamma\left (E\right )$ is ordinary connection expressed in terms of
$E$. There is another polynomial form $[3]$ of the reality conditions.
Perturbatively these two forms are equivalent because both of them ensure
real evolution of the real triad.

Let us choose as a classical background the de Sitter space-time with
the metric
\begin{equation}dS^2=-dt^2+ch^2\left (t\right )d^2\Omega.\end{equation}
$\ d^2\Omega$ is the metric of unit three-sphere. For the sake of simplicity
we put the overall scale factor in $(6)$ to be equal to one. This corresponds
to the cosmological constant $\Lambda=3.$

To find perturbative reduced phase space of the theory, let us decompose
canonical variables into background parts $E$ and $A,$ corresponding to
the metric $(6),$  and fluctuations $H$ and $B,$
\begin{eqnarray*}E\rightarrow E+H,\ A\rightarrow A+B.\end{eqnarray*}

 The linearized constraints take the form
\begin{equation}G_a=D_i{H_a}^i+\varepsilon_{abc}{B_i}^b{E_c}^i=0\end{equation}
\begin{equation}G_i={G_{ij}^a}{E_a}^j+{F_{ij}^a}{H_a}^j=0\end{equation}
\begin{equation}G_0=2\varepsilon^{abc}{H_a}^i{E_b}^j{F_{ij}^c}+\varepsilon^
{abc}{E_a}^i{E_b}^j{G_{ij}^c}-\Lambda\varepsilon^{abc}\varepsilon_{ijk}{H_a}^i
{E_b}^j{E_c}^k=0
\end{equation}

$D_i$ is the background covariant derivative, ${F_{ij}^a}$ is the background
field strength
and ${G_{ij}^a}=\partial_{[i}{B_{j]}}^a+\varepsilon^{abc}{A_{[i}}^b
{B_{j]}}^c.$

We also need linearized gauge transformations of the fluctuations $H$ and
$B.$ In general, an action of infinitesimal gauge transformation generated
by a constraint $G$ on a fluctuation of the variable $Z$ reads
\begin{equation}\delta Z=\left \{\ \int d^3xG\xi\ ,\ Z\right \},\end{equation}
where $\xi$ is the parameter of the transformation and after computation
of the Poisson bracket in the r.h.s. of $(10)$ all the variables should be
replaced by their background values.

In our case the constraints $(2) - (4)$ generate the following
transformations which will be called for short Lorentz, diffeomorphism and
time-evolution respectively.

\begin{displaymath}\;\;\;\;\;\;\;\;\;\;\;\;\;\;\;\;\;\;\;\;\;\;\;\;\;\;\;\;\;
\;\;\;\;\;\;\;\;\;\ \ \ \ \ \ \ \ \delta_L {B_i}^a=iD_i\xi^a
\;\;\;\;\;\;\;\;\;\;\;\;\;\;\;\;\;\;\;\;\;\ \;\;\;\;\;\;\;\;\;\;\;\;\;\;\;\;
\;\ \ \ \ \ (11a)\end{displaymath}

\begin{displaymath}\;\;\;\;\;\;\;\;\ \ \ \ \ \ \ \ \ \ \ \ \ \ \ \ \ \ \ \ \
\ \ \ \ \ \ \ \ \ \ \ \delta_L {H_a}^i=i\varepsilon_{abc}{E_b}^i\xi^c\
\;\;\;\;\;\;\;\;\;\;\;\;\;\ \ \ \ \ \ \ \ \ \ \ \ \ \ \ \ \ \ \ \ \ \ \ \
\ (11b)\end{displaymath}

\begin{displaymath}\;\;\;\;\;\;\;\;\;\;\;\;\;\;\;\;\;\;\;\;\;\;\;\;\;\;
\;\;\;\;\;\;\;\;\;\;\;\;\;\;\;\ \ \delta_D
{B_i}^a=-i{F_{ij}^a}\xi^j\;\;\;\;\;\;\;
\;\;\;\;\;\;\;\;\;\;\;\;\;\;\;\;\;\;\;\;\;\;\;\;\;\;\;\;\;\;\;
\ \ \ (12a)\end{displaymath}

\begin{displaymath}\;\;\;\;\;\;\;\;\;\;\;\;\;\;\;\;\;\;\;\;\;\;\;\;\;\;\;\;\;
\;\;\;\;\ \ \ \ \ \ \ \delta_D {H_a}^i=iD_k\left ({E_a}^{[i}\xi^{k]}\right )
\;\;\;\;\;\;\;\;\;\;\;\;\;\;\;
\ \;\;\;\;\;\;\;\;\;\;\;\;\;\;\ \ \ \ \ \ \ (12b)\end{displaymath}

\begin{displaymath}\;\;\;\;\;\;\;\;\;\;\;\;\;\;\;\;\;\;\;\;
\ \ \ \delta_T {B_i}^a=-2i\xi\varepsilon_{abc}{E_b}^k{F_{ik}^c}-
i\Lambda\xi\varepsilon^{abc}\varepsilon_{ijk}{E_b}^j{E_c}^k\;\;\;\;\;\;
\;\;\;\;\;\;\;\;\;\;\;\;\;\;\ (13a)\end{displaymath}

\begin{displaymath}\;\;\;\;\;\;\;\;\;\;\;\;\;\;\;\;\;\;\;\;\;\;\;\;\;\;\;\;\;
\;\;\;\;\;\delta_T {H_a}^i=iD_k\left (\xi\varepsilon_{abc}{E_b}^{[i}{E_c}^{k]}
\right ),\;\;\;\;\;\;\;\;\;\;\;\;\;\;\;\;\;\;\;\;\;\;\;\;\;\;\;\;\ \
\ (13b)\end{displaymath}

where we used that on the background $(6)$
${F_{ij}^a}=-\varepsilon_{ijk}E^{ak}.$
\setcounter{equation}{13}

One can see that the action of time-evolution transformation on the
the connection fluctuation ${B_i}^a$ is the addition of an arbitrary
proportional to ${E_a}^i$ contribution.
Hence the gauge freedom $(13a,b)$ can be completely fixed by imposing the
following condition on $B$
  \begin{equation}tr\ B\ \equiv{B_i}^a{E_a}^i=0.\end{equation}
 The diffeomorphism transformation  $(12a)$ add an arbitrary antisymmetric term
to the connection $B.$ This part of gauge freedom can be eliminated by the
following condition
\begin{equation}\left (Asym B\right )^{ab}\equiv E^{i[a}{B_i}^{b]}=0.
\end{equation}
After imposing both conditions $(14)$ and $(15)$ the perturbative scalar
constraint $(9)$ is reduced to
\begin{equation}tr\ H\ \equiv{H_a}^i{E_i}^a=0.\end{equation}
Note that in the expression $D_{[i} {B_{j]}}^a$ the Yang-Mills covariant
derivative
$D_i$ can be replaced by totally covariant derivative $\nabla_i$ with
background
Christoffel connection. The background triad $E$ and the tensor $\varepsilon$
commute with this derivative.

The local Lorentz transformation $(11a)$ shifts the connection fluctuation
${B_i}^a$ by a gradient term. Thus a natural gauge fixing condition is
\begin{equation}\nabla^i{B_i}^a=0.\end{equation}

Note that due to the condition $(15)$ the operator $\nabla^i$ in $(17)$ is
real.
However, unlike the previous cases, this gauge freedom can not be
eliminated completely by the last condition. Indeed, undergoing $(14)$
and $(15)$ the $B$ is proportional to a traceless symmetric tensor. The
transformation $(11a)$ of traceless symmetric tensor can be written as
\begin{equation}\delta_L\left (e^{ic}{B_i}^a\right )={i\over 2}
\left (\nabla^a\xi^c+
\nabla^c\xi^a-{2\over 3}\delta^{ac}\nabla_b\xi^b\right ),\end{equation}
where $e^{ic}$ is the unweighted background triad and $\nabla^a\equiv e^{ai}
\nabla_i.$ The operator in the r.h.s. of $(18)$ has zero modes of the form
\begin{equation}{\xi^c}_{(0)}\ \sim\ f^J\left (t\right ) e^{ci} {v_i}^J
\left (x_1,x_2,x_3
\right ),\end{equation}
where ${v^J}_i$ are the ten conformal Killings vectors of $S^3$ and $f^J
\left (t\right )$ are arbitrary functions.

Being imposed the conditions $(14)$ and $(15)$ reduce the vector constraint
$G_i\ (8)$ to
\begin{equation}G_i=\varepsilon_{ijk}{H_a}^jE^{ak}=0.\end{equation}
This means that the matrix ${H_a}^jE^{ak}$ constructed from the triad
fluctuations is symmetric. The Gauss law $(7)$ immediately leads to the
transversality condition on the triad fluctuations
\begin{equation}D_i{H_a}^i=0.\end{equation}
Because of the condition $(20)$ the covariant derivative $D_i$ contains only
real part.

The complex reduced phase space consists of symmetric transverse traceless
fluctuations of the triad and symmetric transverse traceless fluctuations of
the
connection. Let us postpone for a while the discussion on the gauge freedom
$(19)$ which still remains unfixed, and study the reality conditions $(5).$
After linearization they give
\begin{equation}{H^i}_a={H_a}^{i*}\ \quad {B^a}_i+{B_i}^{a*}={1\over 2}
\varepsilon_{abc}
\left (e^{k[b}\nabla_{[i}{h^{c]}}_{k]}+e^{bj}e^{ck}e_{di}\nabla_{[k}{h_{j]}}^d
\right ),\end{equation}
where ${h^c}_k$ is the unweighted triad fluctuation.
We are to verify that these conditions do not destroy the structure of the
reduced phase space. All the conditions $(14)-(17),(20)$ and $(21)$ have the
form of a linear real operator acting on fluctuations. Real and imaginary
parts are restricted independently. Hence the real part of $H$ and the
imaginary part of $B$ satisfy the same symmetry, tracelessness and
transversality conditions. As for the second equation $(22),$ one should
verify that the r.h.s. of this equation will automatically be symmetric,
traceless and transverse. This can be done by straightforward computation.

Consider the gauge freedom $(19).$ It is easy to see that it can not be
eliminated by imposing any condition on variables of the space defined by
$(14)-(17),(20)$ and $(21).$ This situation is similar to that in ADM gravity
$[9]$ and QED $[10]$ on de Sitter space. This gauge freedom can be fixed by
imposing the following condition on Lagrange multiplier $N^a:$
\begin{equation}\int d^3x \ e\ N^a\left (x,t\right )\ {e^i}_a{v_i}^J
\left (x\right )=0\end{equation}
for all ten conformal Killing vectors ${v_i}^J,\quad J=1,\ldots,10.$ Under the
action of Lorentz transformation the variation of $N^a$ looks like following
\begin{equation}\delta N^a=-i\partial_t\xi^a.\end{equation}
Let us find the zero modes of transformation $(24)$ on the space $(19).$ To
this
end let us calculate the scalar product $\langle\delta {N^a}^*,\ \delta N^a
\rangle$
\begin{equation}\langle\delta {N^a}^*,\ \delta N^a\rangle=\int d\tau
\int d^3x\ e\ {\delta N^a
\left (\xi_{(0)}\right )}^*\delta N^a\left (\xi_{(0)}\right ).\end{equation}
This can be done more easily after formal continuation of the integral in the
r.h.s. of $(25)$ to Euclidian space with the $S^4$ metric: $ds^2=d\tau^2+
sin^2\left (\tau\right )d^2\Omega,$ where all relevant operators have discrete
spectrum. Performing integration over $x$ and integration by parts over
$\tau$ and assuming Killing vectors $v^J$ to be orthonormal we obtain
\begin{equation}-\sum_{J}\int d\tau\ sin\left (\tau\right )\left (f^{*J}
\left (\tau\right )\right )
\left [{\partial^2}_\tau+ctg\left (\tau\right )\partial_\tau-{1\over sin^2
\left (\tau\right )}+2\right ]\left (f^{J}\left (\tau\right )\right ).
\end{equation}
$f\left (\tau\right )$ can be expressed as a power series in eigenfunctions
of the self-adjoint operator in $(26).$ These eigenfunctions are just the
associated Legendre polynomials ${P_n}^1\left (cos\left (\tau\right )\right ),
\ n=1,2,\ldots\ .$ The eigenvalues are $-n\left (n+1\right )+2.\ $ We note that
there are ten zero modes $f^J$ corresponding to $n=1$ and $J=1,\ldots,10.$ They
are related to ten Killing vectors of $S^4$ (or de Sitter space in Minkowski
signature). We conclude that the condition $(24)$ give complete fixation of
gauge freedom up to the Killing vectors of the space-time. The Killing vectors
should be anyhow excluded from the gauge group. In covariant gravity they
are excluded from the diffeomorphism group.

In the path integral quantization the gauge freedom $(19)$ can be
incorporated in the following way. Note that the Lagrange multipliers $N^a$
proportional to conformal Killing vectors of $S^3$ do not give any new
constraints. Hence such multipliers should be excluded from the integration
measure by means of $\delta-$function of the conditions $(23).$ A Jacobian
factor should also appear in the integration measure. This procedure was
considered in more details for ADM gravity on de Sitter Space $[6].$
\eject
In conclusion, let us summarize the results of this paper and give several
remarks.

(i) By solving perturbative constraints and fixing (linearized) gauge freedom
on de Sitter background we demonstrated that the complex reduced phase space
consists of symmetric transverse traceless fluctuations of the triad and
symmetric transverse traceless fluctuations of the connection. Our procedure
has
the nice property that the constraints eliminate the same components of the
triad which are excluded in the connection field by the corresponding gauge
fixing. Thus no problem could arise with admissibility of gauge fixing.

(ii) Application of reality conditions do not destroy the structure of the
reduced phase space. In particular, the real part of the connection constructed
from the constrained triad satisfies all gauge conditions.

(iii) A part of gauge freedom can be fixed only by imposing some condition on
Lagrange multiplier.

(iv) It is easy to see that the real reduced phase space of Ashtekar's gravity
is equivalent to that of ADM quantum gravity on de Sitter background
$[6]$.
\vskip .9in
{\bf Aknowledgement}
\vskip .4in
This work was partially supported by the Russian Foundation for Fundamental
Studies, grant 93-02-14378.
\eject
{\bf References.}
\vskip .3cm
\parskip 8pt
\begin{enumerate}
\item
A.Ashtekar, Phys. Rev. Lett. {\bf 57}, 2244 (1986);\\
Phys. Rev. {\bf D36}, 1587 (1987).
\item
A.Ashtekar, Lectures on non-perturbative canonical gravity (World Scientific,
Singapore, 1991)
\item
A.Ashtekar, J.D.Romano, R.S.Tate, Phys. Rev. {\bf D40}, 2572 (1989);\\
I.Bengtsson and P.Peldan, Phys. Lett. {\bf B244}, 261 (1990).
\item
M.Manojlovic and A.Mikovic, Nucl. Phys. {\bf B382}, 148 (1992);\\
Class. Quantum Grav. {\bf 10}, 207 (1993).
\item
H.-J.Matschull and H.Nicolai, Canonical quantum supergravity in
three dimensions, Preprint {\bf DESY 93-073}, gr-qc/9306018;\\
M.Manojlovic and A.Mikovic, Nucl. Phys. {\bf B385}, 571 (1992).
\item
D.V.Vassilevich, Int. J. Mod. Phys. {\bf A8}, 1637 (1993).
\item
R.Arnowitt, S.Deser and C.W.Misner, Phys. Rev. {\bf 116}, 1322 (1959).
\item
L.D.Faddeev, Teor. Mat. Fiz. {\bf 1}, 3 (1969);\\
L.D.Faddeev and A.A.Slavnov, Gauge Fields: Introduction to Quantum Theory
(Benjamin/Cummings, 1980);\\
L.D.Faddeev and R.Jackiw, Phys. Rev. Lett. {\bf 66}, 1692 (1988).
\item
P.A.Griffin and D.A.Kosower, Phys. Lett. {\bf B233}, 295 (1989).
\item
D.V.Vassilevich, Nuovo Cimento {\bf 104A}, 743 (1991); {\bf 105A}, 649
(1992);\\
I.P.Grigentch and D.V.Vassilevich, Nuovo Cimento, {\bf 107A}, 221 (1994).
\end{enumerate}
\end {document}